\documentclass[10pt]{article}
\usepackage{graphicx}
\usepackage{cancel}[]
\usepackage{amssymb,amsmath}
\usepackage{ifpdf}
\usepackage{subfigure}
\usepackage{hyperref}[]
\usepackage{fullpage}
\usepackage{color}
\usepackage{ulem}
\usepackage{verbatim}
\usepackage{float}
\ifpdf
    \DeclareGraphicsExtensions{.pdf}
		
\else
    \DeclareGraphicsExtensions{.eps}
\fi

\usepackage{graphicx,epsfig}
\usepackage{cancel}[]
\usepackage{amssymb,amsmath}
\usepackage{ifpdf}
\usepackage{subfigure}
\usepackage{hyperref}[]
\ifpdf
    \DeclareGraphicsExtensions{.pdf}
\else

\def\lsim{<\kern-2.5ex\lower0.85ex\hbox{$\sim$}\ }
\def\rsim{>\kern-2.5ex\lower0.85ex\hbox{$\sim$}\ }

\def\LAMBDABAR
{\hbox{$\lambda$\kern-0.52em\raise+0.45ex\hbox{--}\kern+0.2em}}

    \DeclareGraphicsExtensions{.eps}
\fi

\begin{document}

 \centerline{\Large\bf{On the Possible Detection of Low Frequency Periodic Signals}}
 \centerline{\Large\bf{in Gravitational Wave Interferometers}}
\vspace{.20in}
 \centerline{Adrian C. Melissinos }
\centerline{\it{Department of Physics and Astronomy, University of Rochester }}
\centerline{\it{Rochester, NY 14627-0171, USA}}
\vspace{.10 in}

\centerline{ 11/10/2014}

\vspace{.2 in}
\vspace{0.5 in}

{\large{\bf{Abstract}}\\

We carried out a computer simulation of a large gravitational wave (GW)
interferometer using the
specifications of the LIGO instruments. We find that if in addition to the
carrier, a single sideband offset from
the carrier by the fsr frequency
(the free spectral range of the arm cavities)
is injected, it is equally sensitive to GW signals as is the
carrier. The amplitude of the fsr sideband signal in the DC region
is generally much less subject to noise than the carrier, and
this makes possible the detection of periodic signals with frequencies well
below the so-called ``seismic wall". The simulation also
explains the observation of tidal gradients, with typical frequencies of
\mbox{$10^{-5}$ Hz,} and strain equivalent $h\sim 10^{-22}$, recorded during the
LIGO S5 run and reported for the LSC in \cite{MG12}.

\section{Introduction}
It is often stated that the sensitivity of gravitational wave interferometers,
as currently operated, is limited to  frequencies
above $\sim 40$ Hz because of ground vibration and other noise in the DC region.
Nevertheless, very low frequency gravity gradients acting on the interferometer
have been extracted from the data. Such observations can be facilitated by examining
the response of a sideband displaced from the carrier by one free spectral range,
$\nu_1 =\nu_0 + \nu_{fsr}$, where $\nu_{fsr} = c/2L$ with $L$ the length of the
interferometer arms. The observable signal is a phase difference due to the travel time
in the two arms, induced by either: (1) the motion of the ``free" end mirrors,
caused by the passage of an appropriately polarized gravitational wave, or
(2) the presence of differential gravity gradients in the two arms which modify the
phase velocity of the propagating light, or (3) geophysical effects that physically
displace the suspension of the end mirrors (test masses). In case (1) we speak
of the ``indirect" coupling of the GW to the
interferometer, while in (2) the gravitational potential (associated with  the
gradient) couples  ``directly"  to the light circulating in the
arms. In GW interferometers, the induced time-dependent phase difference imposes
audio sidebands  on the light circulating in the arms.
The circulating optical field can be either the carrier or the above mentioned $\nu_1$
sideband. The $\nu_1$ sideband has been chosen because it resonates in the arms
and its transfer function is the same as that of the carrier.
This is well known, and indicated in Fig.1 where the transfer function, from end-mirror motion
to the demodulated amplitude at the dark port of the interferometer, is shown  for
both the carrier and the $\nu_1$ sideband, as a function of the excitation frequency.

\begin{figure}[H]
 \centering
 \includegraphics[width=130mm,height=90mm]{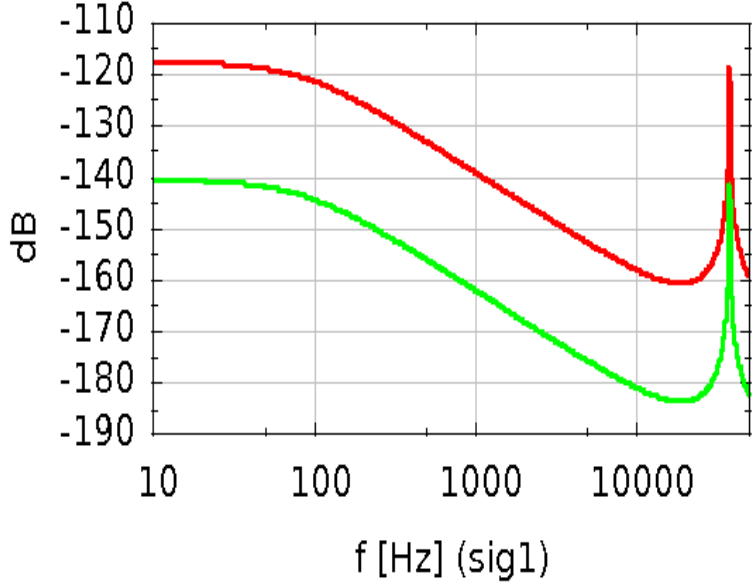}
 \caption{The transfer function, from end-mirror motion
to the demodulated amplitude at the dark port of the interferometer, for
both the carrier (red) and the $\nu_1$ sideband (green), as a function
of excitation frequency. The difference
in scale is due to the different power entering the
interferometer at the carrier and at the sideband frequencies.}
 \end{figure}

The plots were generated with an excitation
of strain $h \sim 10^{-19}$ and as expected, the response is flat below $ f\sim 100$ Hz.
The difference in scale results from  the different power entering the
interferometer at the two frequencies.\\

With the interferometer on a dark fringe, the audio sideband imposed on the carrier,
is detected by mixing (in the photodetector) the carrier with the radio frequency (rf)
sidebands, and then demodulating
the photocurrent at the rf frequency. If the photodetector output is sampled fast
enough, the spectrum of demodulated frequencies will extend beyond $\nu_{fsr}$. The light
at the sideband frequency, $\nu_1$, contains the audio sidebands just as does the carrier,
and the audio sidebands appear in the spectrum, symmetrically displaced about
the $\nu_{fsr}$ line. The audio sidebands can be extracted by demodulating the signal in the fsr region at
the injected $\nu_{fsr}$ frequency. Alternately, when the power in the demodulated signal in the $\nu_{fsr}$
region is plotted as a time series it will be amplitude modulated at the audio frequency.
Spectral analysis of the time series then reveals the frequency and amplitude of the audio
sidebands. The advantage of extracting the audio signal from the $\nu_{fsr}$ amplitude,
rather than from the carrier, is that the low frequency disturbances that dominate the demodulated
carrier signal at $\nu \lesssim 40$ Hz, are suppressed in the fsr region. Of course noise arising
from mirror motion will be present on both the carrier and on the fsr sideband.
However, since the interest is  in low frequency periodic signals, when spectrally analyzing
the time record of the $\nu_{fsr}$  power to identify the audio sidebands, long integration times
can be used, further reducing low frequency random noise.
Signals at frequencies as low as $\nu \sim 10^{-8}$ Hz, have been identified.\\

The demodulated amplitude in the region of the fsr frequency is the sum of the amplitude
due to the fsr sideband $A_{fsr}$ and the audio sideband amplitude $A_{\omega}$ imposed
on it. Thus the power in the fsr frequency region
$$ P = |A_{fsr} + A_{\omega}|^2 = |A_{fsr}|^2 + 2|A_{fsr}||A_{\omega}|{\rm{cos}}(\omega t + \phi) + |A_{\omega}|^2,$$
modulated at the audio sideband frequency $\omega$. The modulation depth, $M$, is
$$ M = \frac{P_{max} - P_{min}}{P_{max} + P_{min}} = 2 \frac{|A_{fsr}||A_{\omega}|}
{|A_{fsr}|^2 + |A_{\omega}|^2}$$
is a measure of the audio amplitude since $|A_{fsr}|$ is fixed and can be directly measured.
For small values of $|A_{\omega}|/|A_{fsr}|$, such as prevailed during the S5 run,
$ M \approx 2|A_{\omega}|/|A_{fsr}|$. In this case,
to detect weak audio amplitudes $A_{\omega}$ it is desirable to keep $A_{fsr}$ as small
as possible.\\

To minimize $A_{fsr}$ at the detection port, the  paths that the fsr sideband $A_{\nu 1}$ follows in
returning from the two arms to the photodiode
 should lead to destructive interference. For the interferometer to be on a dark fringe for the
{\bf{fsr sideband}}, it is not sufficient for the carrier to be on a dark fringe, but it is also
necessary that the macroscopic difference between the two paths be null. The macroscopic
path difference is
$$ \Delta z = \delta l + N \Delta L,$$
where $\delta l$ is the imposed ``Schnupp" asymmetry in the recycling cavity, $\Delta L$ is the
length difference between the arms, typically of order 2 cm for the LIGO interferometers during the S5 run, and $N$
the effective number of round trips in the arms. If $\Delta z$ is different from zero, the fsr sidebands
arriving at the detection port will have a phase difference
$$ \Delta \phi_1= 2\pi \frac{\Delta z}{\lambda_1} = 2\pi\frac{\Delta z}{c}( \nu_0 \pm \nu_{fsr})
= \pm 2\pi\frac{\Delta z}{2L}$$
The last equality follows because, by definition, for the locked interferometer
$2\pi \Delta z/\lambda_0 =0$, modulo $2\pi$, and $\nu_{fsr}= c/2L$ with $L$ the length of the arms.
Reducing $\Delta \phi_1$ to null, is discussed in the following section.\\

Another concern is that if both the upper fsr sideband, $\nu_{1+}$, and the lower one, $\nu_{1-}$,
are present, the audio sidebands imposed on the fsr sidebands cancel at the detection port.
This can be seen in the ``phasor" diagram of Fig.2; here the vectors representing
the fsr sidebands are taken along the real axis and 180$^{\circ}$ out of phase, as is the case when
generated by an electro-optic modulator.  The two induced audio sidebands,
the positive frequency (advanced) and the negative frequency (retarded),
are shown superimposed on both fsr sidebands. Thus, the real part of the phasors always cancel,
and so does the field at the audio frequency.
To study the audio signal imposed on $\nu_1$, it is necessary to inject only {\bf{a single}}
fsr sideband, for instance  $\nu_1 = \nu_0 +\nu_{fsr}$.\\
\begin{figure}[H]
 \centering
 \includegraphics[width=60mm,height=45mm]{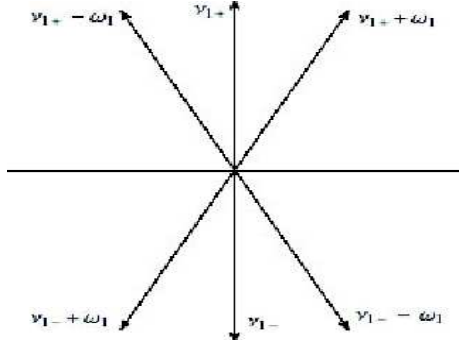}
 \caption{Phasor diagram for the upper $\nu_{1+}$ and lower $\nu_{1-}$
 fsr sidebands when aligned along the real axis, and with the
audio sidebands superimposed. When both fsr sidebands
 are present, the real part of the audio phasor vanishes.}
 \end{figure}
The modulation of the signal at the fsr sideband frequency, $\nu_{fsr}$ was observed by the LIGO
collaboration and has been discussed in the past \cite{MG12, Chad}. Attempts to address the issue
analytically can be found in \cite{Rudenko, Gusev}. However, given the complexity of the optical
fields in GW interferometers, it is advisable to carry out a numerical simulation such as presented here.\\

\section {The numerical simulation}

We have adopted the code FINESSE \cite{Finesse} which uses a standard matrix inversion algorithm
to solve for the fields in the interferometer, as also done in TWIDDLE \cite{Twiddle}.
FINESSE includes a  wide range of useful options and is easy to
use. We simulated a ``standard" recycled interferometer with 4 km long Fabry-Perot cavities
in the arms, using the optical and other parameters of the H1 (Hanford) LIGO interferometer in the
configuration of the S5 run \cite{S5}. The layout of the elements of the interferometer, and the
labeling of the ``nodes" used in the code are shown in Fig.3. We have included rf sidebands
at $\nu_{rf} = 24.480954$ MHz, with modulation index 0.4, and a {\bf{single}} fsr sideband at
$\nu_{fsr} = 37.473$ kHz  (as expected for an arm length of 4,000 m), with modulation index 0.3.
A Schnupp asymmetry of 278 mm (the nominal value for S5) is included.\\

The results of the simulation are shown as graphs generated by the code, either as fixed values
or as  scans over some range of a particular parameter of an element in the interferometer.
In most cases we examine the signals at the detector (AS) port, and plot the demodulated power at
a specific frequency, after demodulating with one, two, or three frequencies. Occasionally, we
give field amplitudes at a particular frequency, or the total power at different points (nodes)
in the interferometer. Power is given in Watts, amplitudes in $\sqrt {\rm{W}}$, and the carrier
input from the laser was set to 1 W.  The end mirrors (or any element) of the interferometer
can be moved sinusoidally (shaken) at a specified frequency and with specified amplitude; the driving amplitude is given
in degrees of phase, namely $\Delta x= \lambda /360 \approx 3\times 10^{-9}$ m, 
which for the end mirrors is equivalent to strain $ h \approx 10^{-12}$.
\begin{figure}[H]
 \centering
 \includegraphics[width=130mm,height=90mm]{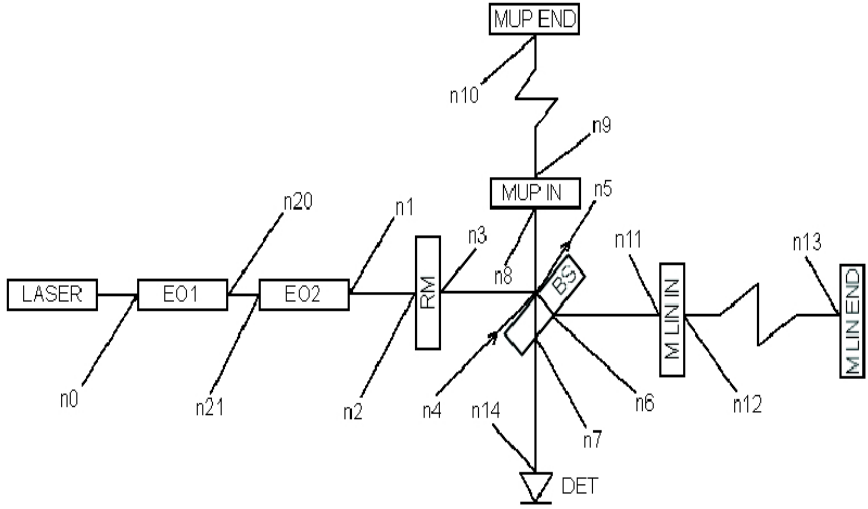}
 \caption{Layout of the elements of the interferometer, and the
labeling of the ``nodes" as used in the simulation code }
 \end{figure}
In the simulation the two end mirrors were shaken differentially. One feature of FINESSE is that
for the carrier the phase at any node is always zero (i.e. microscopically all distances are
taken to be integers of the carrier wavelength), unless specified otherwise by the user
through the addition of a phase at the particular node.\\

The interferometer is set on a dark fringe by choosing the phases of the cavity mirrors appropriately:
for the in-line arm,  input mirror $\phi = 90^{\circ}$, end mirror $\phi = 270^{\circ}$; for the up-going
arm, input mirror $\phi = 0^{\circ}$ and end mirror $\phi = 180^{\circ}$. To place
the fsr sideband ($\nu 1$) on a dark fringe we adjust the length of the in-line arm to
minimize the $\nu 1$ amplitude at the detection port, which happens when
$L_{in-line}$ = 3999.998 m, namely
when the in-line cavity arm is shorter than the up-going arm by 2 mm. This is as expected
because it compensates for the Schnupp asymmetry which is set with the in-line recycling cavity arm
longer than the up-going arm by 278 mm. Taking the effective number of traversals in the arms
as $N \approx 140$, $\Delta z =\delta {\textit{l}} + N\Delta L \approx 0$.
 This is indicated in Fig.4 where the
demodulated fsr power (at the detection port) is shown as a function of the arm-length
of the in-line cavity. We have limited the resolution of the scan to 1 mm, which can be maintained
during interferometer operation without active feedback.\\
  \begin{figure}[H]
 \centering
 \includegraphics[width=130mm,height=90mm]{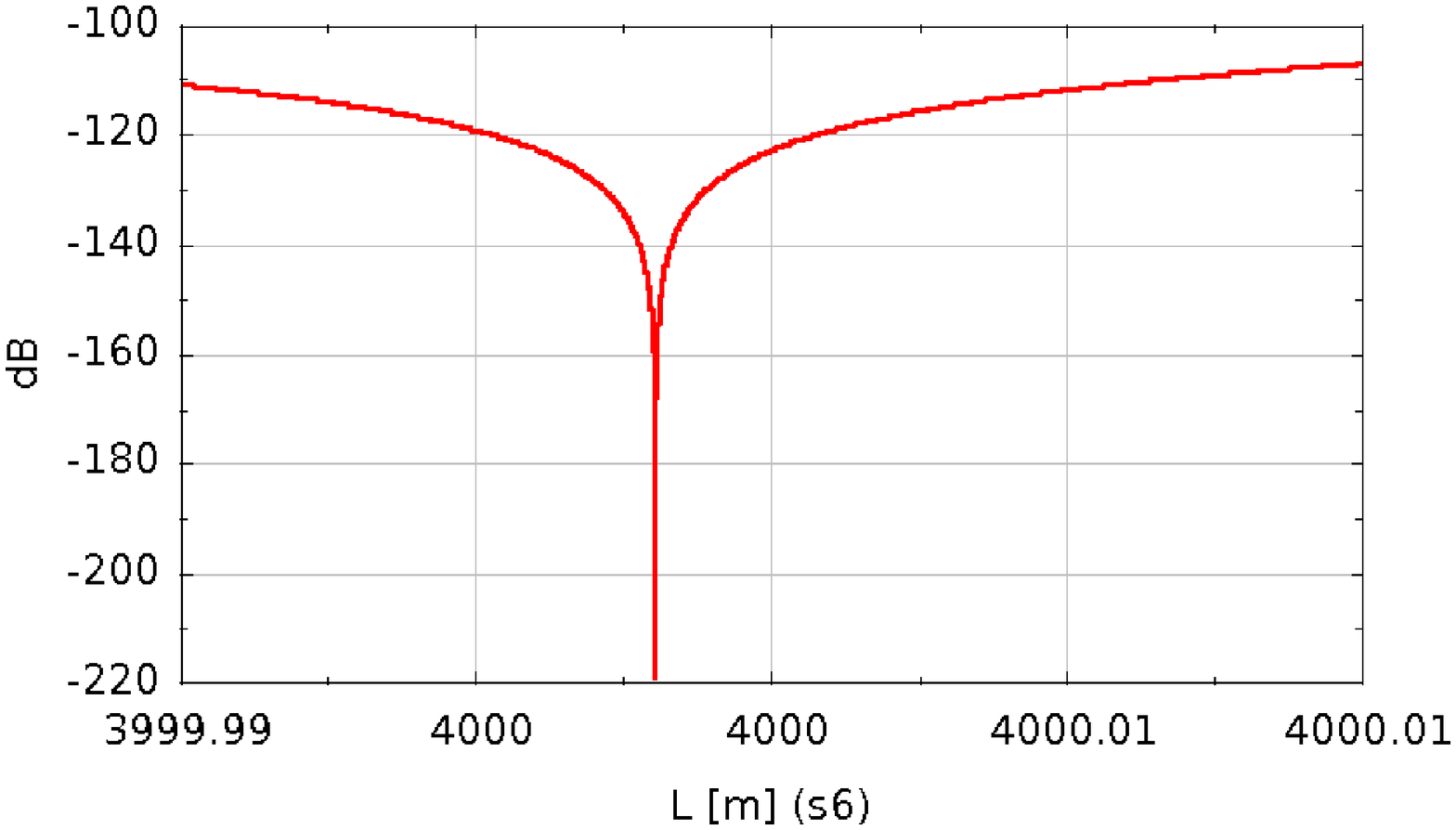}
 \caption{Demodulated power of  the fsr sideband  at the detection
 port as a function of the macroscopic length of the in-line arm}
 \end{figure}
The field amplitudes and demodulated power at the detection port
for this particular tuning  $\Delta L = L_{in} - L_{up} =-2$ mm, are given in Fig.5 . The red line
is the demodulated carrier power, and at -360 db it is indeed dark. The demodulated power of the fsr
sideband is shown by the blue line and at \mbox{ -160 db} it is adequately dark, given the 1 mm resolution chosen
for the setting of the macroscopic arm length difference. The green line shows the
amplitude of the single fsr sideband. These fields are present in the absence of a driving signal,
and with the interferometer in lock. We also checked that the simulation yields the correct power
 gains in the arm cavities, $g^{arm}_0 = 141$ for the carrier, and $g^{arm}_1 = 140$ for the fsr sideband;
in the recycling cavity the carrier gain is $g^{rc}_0 = 111$, and for the fsr sideband $g^{rc}_1 = 100$.\\

 To indicate how the fsr sideband is filtered by the arm cavities and the recycling cavity,
we show in Fig.6 the fsr sideband demodulated power at the detection port as a function of the frequecy
offset from the carrier. The dominant width of the response is due to the spectral width of the
arm cavities, the so called  ``cavity pole". However at the exact fsr frequency, the effect of
the recycling cavity becomes important giving rise to the inverted narrower structure, referred to as
the ``double cavity" pole.\\
\begin{figure}[H]
 \centering
 \includegraphics[width=100mm,height=70mm]{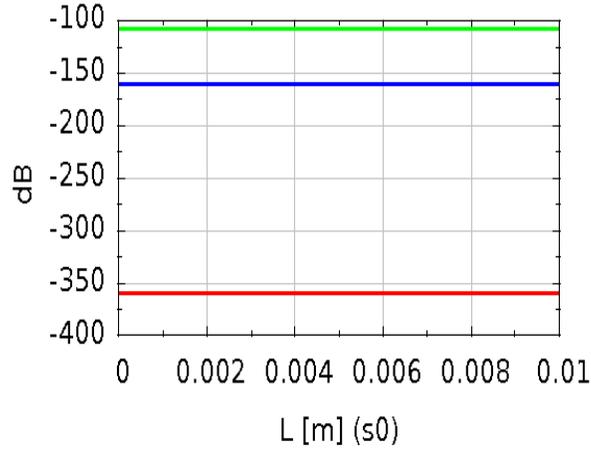}
 \caption{Field amplitude and demodulated power at the detection port
  for optimal tuning and in the absence of excitation. Demodulated carrier
  power (red), demodulated fsr sideband power(blue), amplitude of the
  single fsr sideband (green).}
 \end{figure}

 \begin{figure}[H]
 \centering
 \includegraphics[width=100mm,height=70mm]{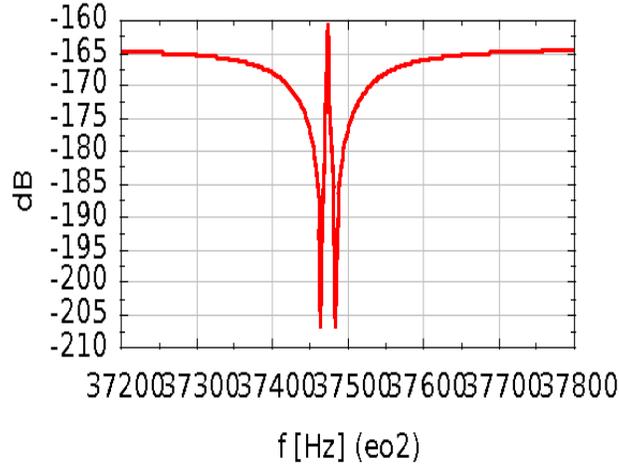}
 \caption{The fsr demodulated power at the detection port as a function of
 the frequency offset (from the carrier) of the injected sideband signal,
 and for ``optimal tuning" of the arm length. See the text for details.}
 \end{figure}
We now turn to the excitation of the interferometer by sinusoidally displacing the end mirror
of both cavities  in a differential manner, that is with opposite phases, as expected
for a suitably polarized GW. We report the ``triply" demodulated power at the detection port.
By this we mean first demodulation at the rf frequency, followed by  demodulation at
the fsr frequency, since we are interested in the audio signal superimposed on the fsr sideband; thirdly
we demodulate at the driving frequency, in order to obtain the response of the interferometer
to the imposed ``shaking" (excitation) of the end mirrors. In Fig.7 the red line gives the response when demodulating
the carrier (the usual signal), while the green line gives the response when demodulating
at the fsr sideband frequency. The horizontal axis is in degrees of phase angle at the end mirrors.
As already mentioned one degree corresponds to a strain $h \approx 10^{-12}$. Thus the horizontal scale spans
$h \sim 10^{-18}\  {\rm{to}}\  10^{-16}$. On this scale the response is linearly dependent
on the excitation amplitude, and the fsr sideband carries the same information about
the excitation as does the carrier. The frequency of the excitation used to generate the signal amplitudes
shown in Fig.7 was 1 Hz, but as indicated in Fig.1
the response is flat below 10 Hz, for both the carrier and the fsr sideband.
The difference in scale between the carrier and the fsr
sideband is determined by the power injected at the two frequencies,
as already mentioned in relation to Fig.1. \\

\begin{figure}[H]
 \centering
 \includegraphics[width=100mm,height=70mm]{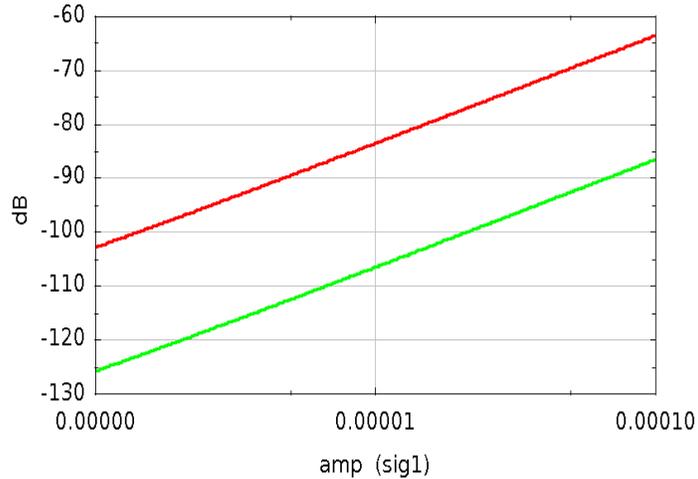}
 \caption{Carrier demodulated power (red) and fsr sideband 
 demodulated power (green) as a
 function of the excitation of the arm end mirrors for the range
 of strain $h \sim 10^{-18}\ {\rm{to}}\ 10^{-16}$. The macroscopic arm length
 difference is set to the optimal value $\Delta L$ = -2 mm.}
 \end{figure}
As a check of our understanding, when both fsr sidebands are allowed in the simulation, the
demodulated fsr power, for the excitation used in Fig.7, is reduced by  200 db  confirming
the conclusion drawn from the  graph in Fig.2.\\

\section{Signal to Noise issues}
To extract the audio signal from the $\nu_1$ amplitude, we follow the approach described in
the introduction, and that was also used in the analysis of the S5 run data:
that is, we examine the modulation of the fsr power as a function of time.
In the introduction we considered the case $A_{\omega} << A_{fsr}$. However when the fsr
amplitude is minimized by adjusting the macroscopic arm length, the
opposite may be true, $A_{\omega} >> A_{fsr}$. The expression for the modulation
depth
$$ M = \frac{P_{max} - P_{min}}{P_{max} + P_{min}} = 2 \frac{|A_{fsr}||A_{\omega}|}
{|A_{fsr}|^2 + |A_{\omega}|^2}$$
is valid in either case.
When $A_{\omega} >> A_{fsr}$, the signal at the audio frequency dominates and it is
directly available. In Fig.8 we show the audio
amplitude (green), the fsr amplitude (red) and the modulation depth, $M$ (blue) for the same
range of excitation amplitudes as used in Fig.7,  namely in the range
$h \sim 10^{-18}\  {\rm{to}}\  10^{-16}$ and at $\nu_{audio} =1$ Hz.  In the simulation, the 
power at the fsr, is
obtained by demodulating the photocurrent, at the detection port, at the rf frequency (24.480954 MHz),
followed by  demodulation at the fsr frequency (37.473 kHz). The power at the audio amplitude is
obtained by following the same procedure as above, and then demodulating for a third time at the 
audio frequency used in the code to drive the end mirrors.
The resulting $A_{fsr}$ amplitude, is obviously independent of the external
drive, while the power at $A_{\omega}$  grows linearly with increasing excitation. 
When the two amplitudes are equal there is 100 percent modulation, $M=1$.\\
\begin{figure}[H]
 \centering
 \includegraphics[width=100mm,height=65mm]{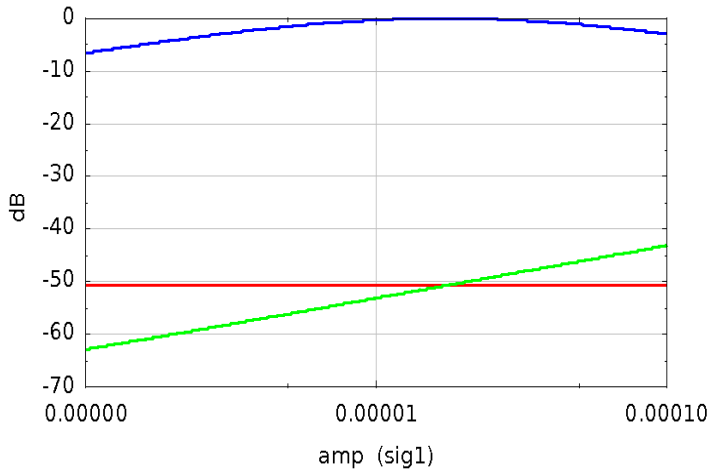}
 \caption{ fsr sideband amplitude (red), audio amplitude (green), and
 modulation depth (blue), as a function of the excitation of the arm
 end mirrors for the range of strain $h \sim 10^{-18}\ {\rm{to}}\ 10^{-16}$.
 When the amplitudes are equal, $M=1$.
 The macroscopic arm length
 difference is set to the optimal value \mbox{$\Delta L$ = -2 mm.}}
 \end{figure}
The FINESSE code calculates the ``noise over signal" ratio, N/S, (or its inverse)
at any node in the interferometer by applying
the Schottky formula to the total optical power at the node and comparing the associated noise 
power to the
signal power at the particular frequency of interest. We have adopted the convention
of presenting ``noise over signal", N/S, which is a spectral density and must
be multiplied by $\sqrt {\rm{BW}}$ with BW the bandwidth used in the measurement,
expressed in Hz. For the results presented here we use
BW=1 Hz. Once the N/S is known for  $A_{\omega}$ and $A_{fsr}$ the errors can be
propagated to obtain the N/S for the modulation depth as well.\\

 In these estimations we consider
{\bf{only shot noise}}, and this is justified because the photocurrent which is demodulated to
yield the power at $\nu = 37.473$ kHz is mainly free of the disturbances near DC.
Some of these disturbances are up-converted from the DC region   to the fsr frequency region,
but the up-converted amplitudes are typically few percent of $A_{fsr}$ \cite{Butler}.
Up-converted discrete low frequency lines, such as due to the suspension of the
optics, can be separated from the audio signal.  Random (white) noise, is further suppressed when
extracting the spectrum of periodic signals from a long time series of the  power in the fsr
frequency range: during the S5 LIGO run the time record was 16 months long. \\
\begin{figure}[H]
 \centering
 \includegraphics[width=100mm,height=70mm]{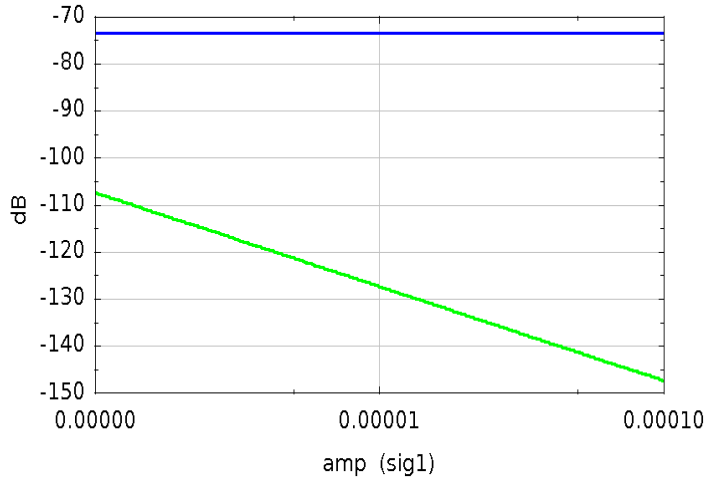}
 \caption{ N/S for the audio demodulated power (green), N/S  for the 
 demodulated power at the fsr sideband
  (red), and N/S for the modulation depth (blue), as a function
 of the excitation of the arm end mirrors for the range of strain
 $h \sim 10^{-18}\ {\rm{to}}\ 10^{-16}$. The macroscopic arm length
 difference is set to the optimal value $\Delta L$ = -2 mm.
 As explained in the text, the red and blue lines overlap.}
 \end{figure}
In Fig.9 we plot the N/S for the demodulated power at he audio frequency (green), 
the N/S for the power at the fsr sideband(red), and the N/S for the modulation 
depth (blue), for the same range of signal excitation as in
Figs.7,8, namely for strain in the range $h \sim 10^{-18}\  {\rm{to}}\  10^{-16}$. 
The red curve is not seen in Fig.9
because it is overwritten by the blue N/S line of the modulation depth.
This is to be expected since for optimal tuning,  $A_{fsr} << A_{\omega}$,
and the N/S for $A_{fsr}$ dominates the N/S for the modulation depth. For optimal
tuning of the interferometer, and for  strain  $ h=10^{-23}$ the signal
to noise ratio, due only to shot noise, is $S/N \approx 4$ for the signal extracted 
from the sideband, while it is $S/N \approx 40$ for the signal extracted from the 
carrier.\\

\section{Discussion}

The response of the detected fsr power to an external excitation was observed serendipitously during
the LIGO S5 run. In that run the fsr amplitude was recorded and  demodulated as a separate
channel in the detection chain, to search for a high frequency gravitational signal. Instead,
the analysis of the fsr data revealed the slow modulation of the time
series of the fsr power at a frequency of $\sim 10^{-5}$ Hz. In the S5 run there
was no injection at the fsr sideband, but the sideband was spontaneously excited by parametric
conversion from a nearby thermal (acoustic) resonance in the  test masses (mirrors).
The fsr sideband as well as the acoustic resonances can be seen in the first figure of \cite{MG12}.
The time series of the power
in the fsr channel (integrated over the line width) for a 16-month long record is plotted in Fig.10;
as seen in the inset it is modulated with a period of half a day and of one day.
In retrospect, the presence of modulation indicates that only one, and not both sidebands
are generated in the parametric conversion process\footnote{This can be attributed to the fact
that the frequency of the acoustic resonance is slightly higher than the fsr frequency.
Thus, preferentially only the upper sideband participates in the parametric conversion.}.
 Spectral (Fourier) analysis
of the time record reveals 10 tidal lines, the observed frequencies being compared with
their known values \cite{Melchior} in Table 1. \\

The spectra of the integrated power at the fsr are shown in Fig.11 for the region of
diurnal frequencies and in Fig.12 for the semi-diurnal region.
 \begin{figure}[H]
 \centering
 \includegraphics[width=100mm,height=70mm]{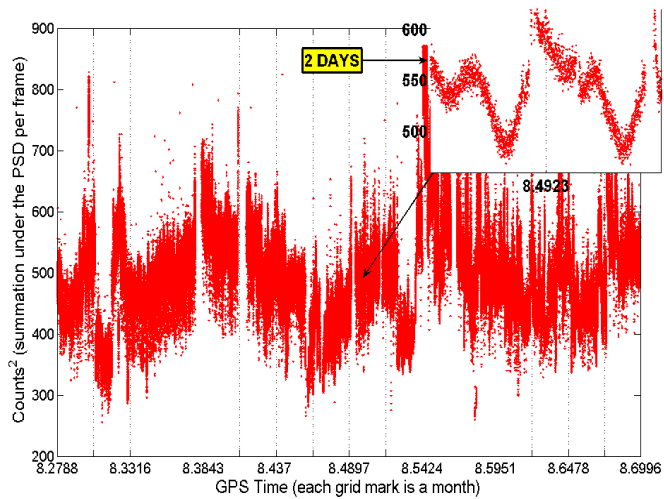}
\caption{Integrated power in the free spectral range (fsr) region as  a
function of time, April 2006 to July 2007. The data are for the H1 interferometer
and are sampled every \mbox{64 s.} Note the daily and  twice-daily modulation
that can be seen in the inset. From \cite{MG12}.}
 \end{figure}
\pagebreak

{\underline { Table I. \ \ Observed and known frequencies of the tidal
components (Hz)}}

\begin{tabular}{c l|l|l} \hline

\qquad Symbol \qquad \qquad  & Measured \qquad \qquad & Predicted \qquad \qquad & L=lunar; S=solar\\
\hline\\[0.2ex]
{\underline {Long period }}\\
 $\rm{Ss_a}$ & $6.536\times 10^{-8}$ & $6.338\times 10^{-8}$ & \rm{S declinational}\\[0.5ex]
 {\underline {Diurnal }}\\
 $\rm{O_1}$ & $1.07601\times 10^{-5}$ & $1.07585\times 10^{-5}$ & \rm{L principal lunar wave}\\
 $\rm{P_1}$ & $1.15384\times 10^{-5}$ & $1.15424\times 10^{-5}$ & \rm{S solar principal wave}\\
 $\rm{S_1}$ & $1.15741\times 10^{-5}$ & $1.15741\times 10^{-5}$ & \rm{S elliptic wave of} $\rm{^{s}K_1}$\\
 $\rm{^{m}K_1,^{s}K_1}$ & $1.16216\times 10^{-5}$ & $1.16058\times 10^{-5}$ & \rm{L,S declinational waves}\\
 [0.5ex]

{\underline {Twice-daily }}\\
 $\rm{N_2}$ & $2.19240\times 10^{-5}$ & $2.19442\times 10^{-5}$ & \rm{L major elliptic wave of $\rm{M_2}$}\\
 $\rm{M_2}$ & $2.23639\times 10^{-5}$ & $2.23643\times 10^{-5}$ & \rm{L principal wave}\\
 $\rm{S_2}$ & $2.31482\times 10^{-5}$ & $2.31481\times 10^{-5}$ & \rm{S principal wave}\\
 $\rm{^{m}K_2,^{s}K_2}$ & $2.31957\times 10^{-5}$ & $2.32115\times 10^{-5}$ & \rm{L,S declinational waves}\\

\end{tabular}\\

As seen in the Table the measured tidal frequencies are in excellent agreement 
with their predicted value,
within the resolution of the measurement. The uncertainty
in the determination of the tidal frequencies is \mbox{$\Delta
\nu_{res}= 1/(4T_{total}) = 6 \times 10^{-9}$ Hz}, with $T_{total} = 4.2
\times 10^7$ seconds; the factor of 4 being included because in the
spectral analysis \cite{Lomb-Scargle} the data was oversampled by a factor of four.
Comparing the observed frequencies to the predicted ones, and using
$\Delta \nu_{res}$ as  the measurement
error, yields $\chi^2$/DF = 1.86. The table also includes a long-term,
twice yearly, component which is evident by inspection of Fig.10.\\

The presence of the Earth tides is well known, and 
to keep the interferometer in lock, the end test masses must be mechanically
displaced to correct for the tidal motion. 
Any residual uncompensated motion is corrected by the
interferometer controls and can be observed in a
long term analysis of the trends of the differential arm control signal (DARM-CTRL). However,
the tidal acceleration has also a horizontal component along the arms, typically
$g_{hor} \approx 10^{-7} g \approx 10^{-6} \ {\mathrm{m\  s^{-2}}}$, and this component is time-dependent
at the tidal frequencies. This horizontal gravity gradient leads to a frequency shift of the light
propagating along the arms, and thus to a cumulative phase shift for every traversal. In the weak field approximation,
the presence of a gravitational potential $\Phi$ modifies the $g_{00}$ metric coefficient to
\begin{equation}
 g_{00} = -(1 + 2\Phi/c^2)
\end{equation}

 \noindent The departure of $g_{00}$ from its flat space value gives
 rise to time dilation, or equivalently to a shift in the frequency
 of light propagating through that gravitational field \cite{ Weinberg,
 Hartle}.
 \begin{equation}
 \nu_A - \nu_B = - \frac{\Phi_A - \Phi_B}{c^2}\ \nu_{A}  \qquad\qquad
 {\rm{or}}\qquad\qquad  \frac{\delta \nu}{\nu} = -\frac{\delta n}{n}
 = -\frac{\delta \Phi}{c^2},
 \end{equation}
 where we also introduced the refractive index of the light $n =
 c'/c$, which is often used in the literature. \\

  A constant gradient
 $g_{hor}$ along the $x$-direction can be described by a potential,
 $\Phi = g_{hor}x$. Thus light executing a single round trip in an arm of
 length $L$ acquires a phase shift (as compared to light traveling in
 a field-free region) equal to
 \begin{equation}
 \delta \phi_t^{single} = 2\int \delta \omega dt = 4\pi\nu_0 \int^{L}_{0}
  \frac{\delta \nu}{\nu} \frac{dx}{c} =
  \frac{4\pi}{\lambda_0}
  \int^{L}_{0} \frac{\Phi}{c^2} dx
 = \frac{2\pi}{\lambda_0}\ \frac{ g_{hor} L^2}{c^2}.
 \end{equation}

 \begin{figure}[H]
 \centering
 \includegraphics[width=100mm,height=70mm]{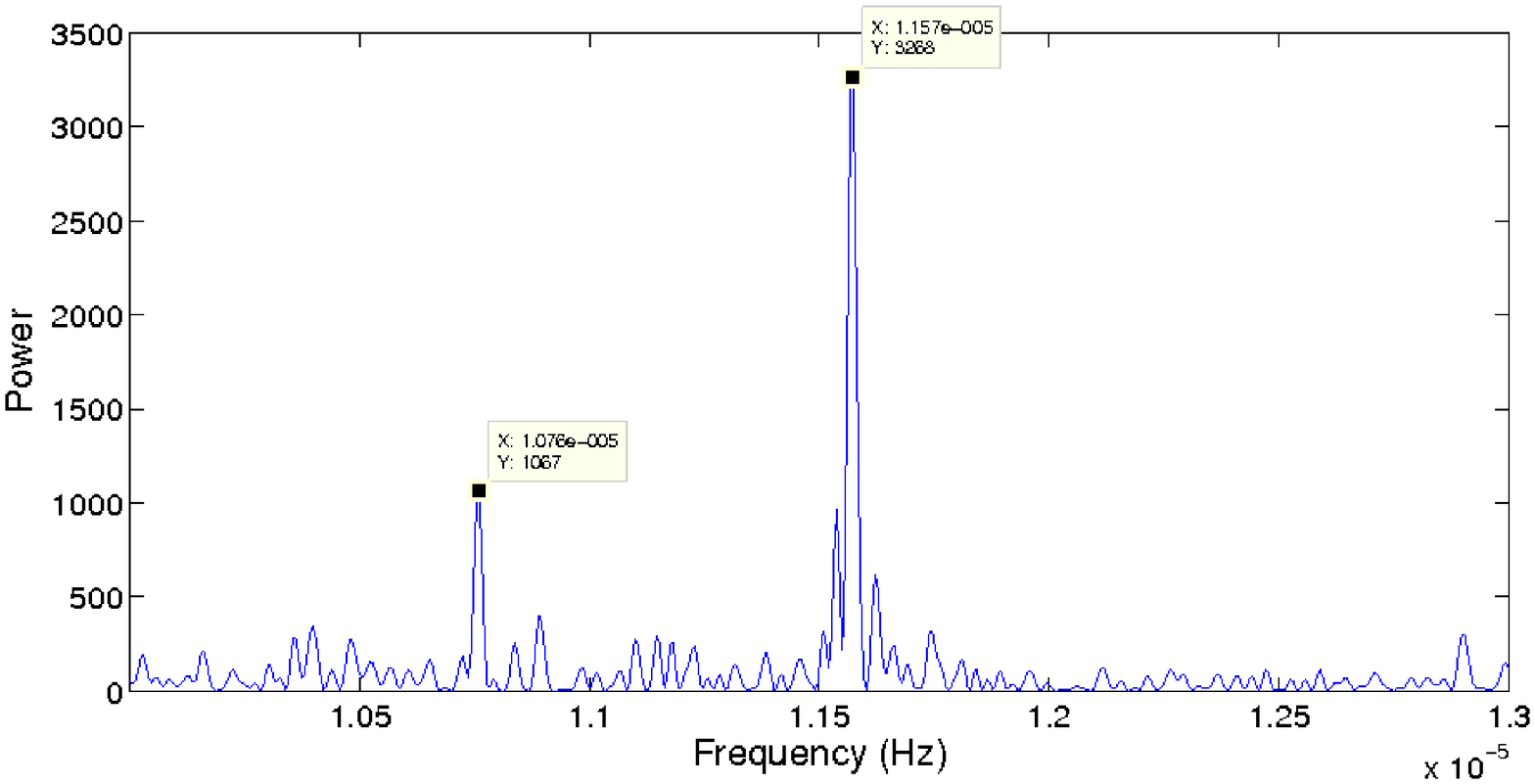}
 \caption{Frequency spectrum of the integrated fsr power in the
 diurnal region, from \cite{MG12}. Note the fine structure.}
 \end{figure}

 \begin{figure}[H]
 \centering
 \includegraphics[width=100mm,height=70mm]{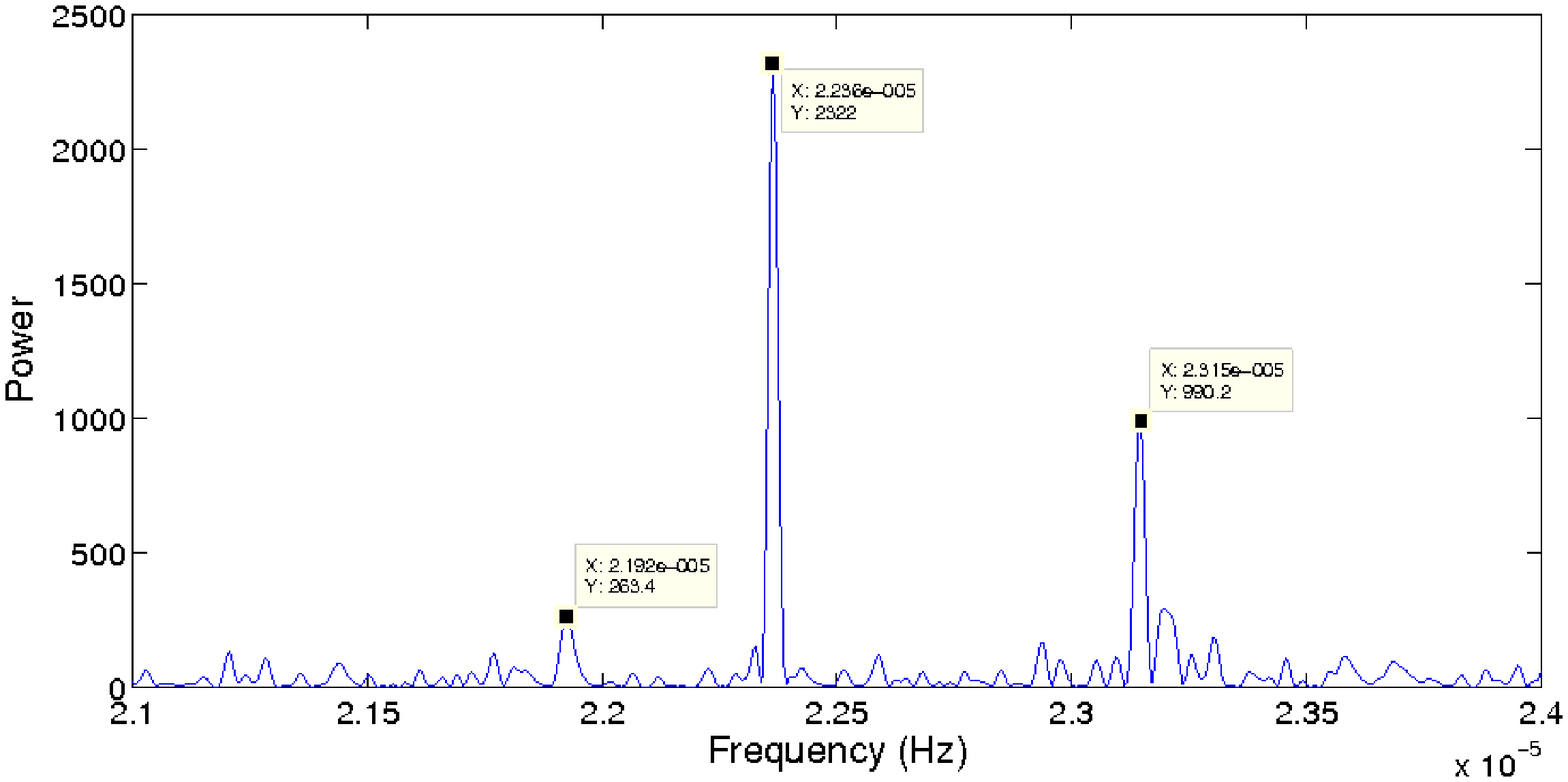}
 \caption{Frequency spectrum of the integrated fsr power in the
 twice daily region, from \cite{MG12}. Note the fine structure.}
 \end{figure}

\noindent Numerically, and without accounting for the multiple traversals, we find
 $$\frac{\Delta \phi}{2 \pi} \sim 2\times 10^{-10}.$$
We can use this value to estimate the modulation depth expected for the LIGO S5
data. To this end we ran the simulation with a macroscopic arm length difference
of $\Delta L =2$ cm (best estimate for the S5 run \cite{Butler}), rather than the ``optimal" value
($\Delta L =- 2$ mm) used in the simulations discussed so far. By introducing an excitation
of $7\times 10^{-8}$ degrees, corresponding to the above value of $\Delta \phi$,
we find a modulation depth of $M=0.18$. This is in qualitative agreement with the data, and
is additional evidence that the observed signals are due to the gravity gradient
and not to uncompensated mirror motion. The advantage of using the modulation depth
as a measure of the induced phase shift, is that it does not depend on the power input
at the fsr sideband,
since both the $A_{fsr}$ and $A_{\omega}$ amplitudes scale by the same factor.\\

In previous attempts to explain the modulation of the fsr power, macroscopic mirror motion was
often considered \cite{Chad, Rudenko, Gusev}. Such motion, however, is compensated by the tidal servo and any remaining
displacement is corrected by the interferometer controls. Instead, we now see
that any external audio excitation, such as produced by the gravity
gradients will simultaneously impose audio sidebands on both the carrier and on the fsr
amplitude. The information carried by the fsr amplitude is the same as  that
carried by the carrier, except that the audio signals are now superimposed on a
higher frequency signal which is displaced from the DC region where seismic noise and
similar disturbances are dominant. The tidal signals have also been extracted from
the minute trends of the signal controlling the differential arm separation
(DARM-CTRL) but in that case the frequencies of the dominant lines are
shifted by $d\nu/\nu \approx 0.005$ and the weaker lines are absent. This may be due to the higher
noise level when the low frequency signals are extracted from the carrier. Of course,
the demodulation in the DC region
performed on the carrier, is essential for sensing low frequency motion in order to keep the
interferometer in lock. But once the interferometer is locked,  when searching for low frequency
audio signals, it is best to examine the demodulated power in the fsr amplitude.

\section {G.W. signals from close binaries}
As seen in Figs.11,12 the LIGO interferometers can detect weak signals at low frequencies
$\nu \sim 10^{-5}$ Hz, with excellent signal to noise ratio (SNR) .
While the horizontal tidal gradient is large, $\sim 10^{-6}\ {\rm m s^{-2}}$,  it
is coupled to the interferometer through the ``direct" effect imposing only a very
small phase shift.  An incoming GW  (at a sufficiently high frequency)
that would produce the same phase shift by acting on the ``free"
mirrors, would have a strain $h=(\Delta \phi/2 \pi)(\lambda/ NL) \sim 5\times 10^{-22}$. Therefore
it may be possible to detect the gravitational signal from close binaries that have typical
periods of a fraction of a day or less. \\

The main difficulty is that at frequencies below the natural frequency
of the suspension the test masses can not any more be treated as free.
But not all is lost: the test masses respond to the tidal acceleration imposed by
the gravitational wave \cite{MTW, Rakhmanov}
$$ \ddot x = \frac{1}{2} \ddot h(t) x$$
which is counteracted by the restoring acceleration of the pendular suspension.
Letting $\Omega, h$ be the angular frequency and amplitude of the gravitational wave,
and $\omega_s$ the pendular angular frequency, a direct calculation for the
change in the arm length gives\footnote{The following equation has the inverse behavior 
from that of the driven
oscillator because the driving acceleration is proportional to $\Omega^2$.}
\begin{equation}
\Delta L(t) = -\frac{h_0 L}{2} \frac{1}{(\omega_s/\Omega)^2 - 1}e^{i \Omega t}
\end{equation}
as compared to $\Delta L (t) = (h_0 L/2)e^{i \Omega t}$ when $\Omega >> \omega_s$, namely 
when the mirrors are free.
We see that in calculating the phase shift in the interferometer, the 
strain of a low frequency gravitational wave must be derated\footnote{The direct coupling is always present,
but the induced phase shift  $\Delta \phi = (k h LN)(\Omega T)^2/6$, scales
as the square of the g.w. frequency.} by $(\Omega /\omega_s)^2$.\\

For the majority of binaries, $\Omega/2 \pi \sim 10^{-5}$ Hz, while for LIGO
$\omega_s = 0.75$ Hz and for VIRGO $\omega_s = 0.5$ Hz. Given the expected gravitational
wave strain produced by binaries, a large improvement  in interferometer sensitivity
appears necessary to detect such sources. However, Brown et al. reported recently on a pair of detached white
dwarfs with a period of $T=12.75$ minutes ($\Omega /2 \pi =2.6 \times10^{-3}$ Hz)
\cite{Brown}. The binary is located at a distance of 1 kpc, and the calculated
gravitational wave strain at the Earth is $h = 10^{-22}$. To detect the source in
one year of observation, the required sensitivity of the interferometer is
$$ h/\sqrt {\rm{Hz}} = 10^{-22} (\Omega/\omega_s)^2
(6 \times 10^{-9})^{-1/2}
\sim 4\times 10^{-23}\sqrt {\rm{Hz}}$$
for $\omega_s = 0.5$ Hz. This seems achievable, and there may exist other binaries
with even shorter periods, that have not as yet been detected optically but
could be searched for, through their gravitational signal.

\section{Acknowledgements}
I wish to thank Daniel Sigg who designed and implemented the fsr read-out
channel, and Fred Raab for his support and hospitality at the Hanford
LIGO laboratory. Bill Butler, Chad Forrest, Tobin Fricke and Stefanos Giampanis
were instrumental in the analysis. The excellent quality of the data
is due to the dedicated efforts of the staff and operators at LHO, and of
the members of the LSC collaboration. I also thank Valentin Rudenko
of Moscow State University, for his encouragement and continuing
interest in this work.  I am indebted to the team that wrote and
maintains the FINESSE code, and I thank Mark Bocko and Malik Rakhmanov for
insightful comments on an earlier version of this paper.

\end{document}